\documentclass[reqno]{amsart}
\usepackage{amssymb}

\newcommand{\eq}[1]{\eqref{#1}}
\newcommand{\up}[1]{^{(#1)}}

\numberwithin{equation}{section}
\newtheorem{theorem}{Theorem}[section]

\newtheorem{lemma}[theorem]{Lemma}
\newtheorem{corollary}[theorem]{Corollary}

\newtheorem{remark}[theorem]{Remark}

\newenvironment{acknowledgement}{\emph{Acknowledgement.}}

\DeclareMathOperator{\supp}{supp}
\DeclareMathOperator{\tr}{tr}

\newcommand\beq{\begin{equation}}
\newcommand\eeq{\end{equation}}

\renewcommand\H{\mathcal{H}}

\newcommand\R{\mathbb R}
\newcommand\N{\mathbb N}
\newcommand\C{\mathbb C}
\newcommand\Z{\mathbb Z}

\newcommand\e{\mathrm{e}}
\newcommand{\la}{\langle}
\newcommand{\ra}{\rangle}
\renewcommand\P{\mathbb P}
\newcommand\E{\mathbb E}

\renewcommand{\d}{\mathrm{d}}
\newcommand\vphi{\varphi}

\newcommand\eps{\varepsilon}

\newcommand{\bom}{{\boldsymbol{\omega}}}
\newcommand{\btau}{{\boldsymbol{\tau}}}
\newcommand{\bmu}{{\boldsymbol{\mu}}}

\newcommand{\abs}[1]{\left\lvert #1 \right\rvert}
\newcommand{\norm}[1]{\left\lVert #1 \right\rVert}
\newcommand{\scal}[1]{\left\langle #1 \right\rangle}
\newcommand{\set}[1]{\left\{ #1 \right\}}
\newcommand{\pa}[1]{\left( #1 \right)}
\newcommand{\hnorm}[1]{\left\{ \!\left\{ #1\right\}\! \right\}}

\begin{document}

\title[Generalized eigenvalue-counting estimates]{Generalized eigenvalue-counting estimates for the Anderson model}

\author[J.-M. Combes]{Jean-Michel Combes}
\address[Combes]{Universit\'e du Sud: Toulon et le Var, D\'epartement de Math\'ematiques,
F-83130 La Garde, France}
\email{combes@cpt.univ-mrs.fr}

\author[F. Germinet]{Fran\c cois Germinet}
\address[Germinet]{Universit\'e de Cergy-Pontoise,
CNRS UMR 8088, IUF, D\'epartement de Math\'ematiques,
F-95000 Cergy-Pontoise, France}
\email{germinet@math.u-cergy.fr}

\author[A. Klein]{Abel Klein}
\address[Klein]{University of California, Irvine,
Department of Mathematics,
Irvine, CA 92697-3875,  USA}
 \email{aklein@uci.edu}

\thanks{A.K was  supported in part by NSF Grant DMS-0457474.}

%\date{Version of \today}
%%%%%%%%%%%%%%%%%%%%%%%%%%%%%%%%%%%%%%%%%%%%%%%%%%%%

\begin{abstract}
We  generalize Minami's estimate for the Anderson model and its extensions to $n$ eigenvalues, allowing for  $n$ arbitrary intervals and arbitrary single-site probability measures with no atoms.    As an application, we derive  new results about the multiplicity of eigenvalues and Mott's formula for the ac-conductivity when the single site probability
distribution is H\"older continuous.
\end{abstract}

%%%%%%%%%%%%%%%%%%%%%%%%%%%%%%%%%%%%%%%%%%%%%%%%%%%%
\maketitle

%%%%%%%%%%%%%%%%%%%%%%%%%%%%%%%%%%%%%%%%%%%%%%%%%%%%%%%%%%%%%%%%%%%%%%%%%%%%%%%%%
%%%%%%%%%%%%%%%%%%%%%%%%%%%%%%%%%%%%%%%%%%%%%%%%%%%%%%%%%%%%%%%%%%%%%%%%%%%%%%%%%
%%%%%%%%%%%%%%%%%%%%%%%%%%%%%%%%%%%%%%%%%%%%%%%%%%%%%%%%%%%%%%%%%%%%%%%%%%%%%%%%%

\section{Introduction}

We consider the generalized Anderson model  given by the random Hamiltonian
\begin{equation}
H_{\bom} = H_0 + V_{\bom} \quad \text{on}\quad \ell^2(\Z^d),\label{defH}
\end{equation}
where $H_0$ is a  self-adjoint operator and $V_{\bom}$ is the random potential
given by $V_{\bom}(j)= \omega_j$.  Here
 $\bom=\{ \omega_j \}_{ j\in
\Z^d}$ is a family of independent
random
variables;  $\mu_j $ will denote the probability distribution of  the random variable $\omega_j $.  In this article we always assume that each $\mu_j$ has no  atoms.    We write $\E_{\omega_j}$ for the expectation with respect to the random
variable $\omega_j$, and  write $\E=\E_\bom$ for the joint expectation.  We also set
$\bom^\perp_k=\{ \omega_j \}_{ j\in
\Z^d\setminus \set{k}}$ and let $\E_{\bom^\perp_k}$ denote the corresponding expectation.

Restrictions of $H_\bom$ to finite  volumes $\Lambda\subset\Z^d$  are denoted by
$H_{\bom,\Lambda}$, a self-adjoint operator of the form
\begin{equation}
H_{\bom,\Lambda } = H_{0,\Lambda} + V_{\bom,\Lambda } \quad
\text{on}\quad \ell^2(\Lambda),\label{defHfinite}
\end{equation}
with $H_{0,\Lambda}$  a self-adjoint restriction of $H_0$ to the finite-dimensional
Hilbert space  $ \ell^2(\Lambda)$ and $ V_{\bom,\Lambda }(j)=\omega_j$ for $j \in \Lambda$.
(The results discussed in this article are not sensitive to the choice of $H_{0,\Lambda}$.)    Given a Borel set $J \subset \R$, we write
$P^{(\Lambda)}_{\bom}(J)=P^{(\Lambda)}_{H_\bom}(J)=\chi_J(H_{\bom,\Lambda})$ for  the
associated spectral projection.

Minami \cite{M} estimated the probability that  $H_{\bom,\Lambda}$ has at least two
eigenvalues in an interval $I$.  Assuming that all $\mu_j$ have bounded densities
$\rho_j$,
Minami proved that
\begin{equation}\label{minamiest}
2\,\P\set{\tr P^{(\Lambda)}_{H_\omega}(I)\ge 2 }\le \E  \set{\pa{\tr
P^{(\Lambda)}_{H_\omega}(I)}^2 -\tr P^{(\Lambda)}_{H_\omega}(I) } \le \pa{\pi
\rho^{(\Lambda)}_\infty \abs{I} \abs{\Lambda}}^2,
\end{equation}
where $\rho^{(\Lambda)}_\infty:= \max_{j\in \Lambda} \norm{\rho_j}_\infty$.  Minami's
proof required  $H_0$ to have  real matrix elements, i.e., $ \langle \delta_j,  H_0
\delta_k \rangle \in \R$ for all $j,k $. This restriction was recently  removed by
Bellissard,  Hislop and Stolz \cite{BHS} and by   Graf and Vaghi \cite{GV}.  They also
estimated the probability that  $H_{\bom,\Lambda}$ has at least $n$ eigenvalues in  $I$
for all $n \in \N$,  assuming, as Minami,  that all $\mu_j$ have bounded densities
$\rho_j$,

Minami's estimate has important consequences for the physical behavior of the  Anderson
model in the localized (insulator) regime. It is the crucial ingredient in Minami's proof
of the absence of eigenvalue repulsion, showing that the properly rescaled eigenvalues
behave according to a Poisson process \cite{M}. (See \cite{N,KN,Kr,St1,St2} for further
developments.)  It was shown to imply simplicity of  eigenvalues by Klein and Molchanov
\cite{KM}. It is an important ingredient in the derivation of a rigorous form of Mott's
formula for the ac-conductivity by Klein, Lenoble and M\"uller \cite{KLM}.

 In \cite{CGK} we introduced a new approach to Minami's estimate.
The  crucial step in Minami's proof, namely \cite[Lemma~2]{M},   estimates the average of  a determinant whose entries are matrix elements of the imaginary part of the resolvent; the proofs
in \cite{BHS,GV} have similar steps.    In contrast,
 our method only averages  spectral projections, which allowed us to  finally prove a Minami estimate for  the continuum Anderson Hamiltonian. As a consequence, we obtained
  Poisson eigenvalue statistics and simplicity of eigenvalues  for the continuum Anderson Hamiltonian.
  
 The  new approach, in addition to providing a   simple and transparent proof  of Minami's estimate for the Anderson model, also allows for arbitrary single-site probability measures with no atoms.        Given a probability measure $\mu$, we let $S_\mu(s):=\protect{ \sup_{a \in \R}
\mu([a,a+s])}$, 
  the concentration function of $\mu$, and set 
   \beq \label{defQ}
Q_\mu(s):=
\left\{ 
\begin{array}{ll}
\norm{\rho}_\infty s &\mbox{if $\mu$ has a bounded density $\rho$} 
\\
 8 S_\mu(s) &\mbox{otherwise}
\end{array}.
\right.\notag
\eeq
 (Note that the measure $\mu$ has no atoms if and only if $\lim_{s
\downarrow 0} Q_\mu(s)=0$.)
For the generalized Anderson model $H_\bom$  as in \eq{defH}, we let $Q_j=Q_{\mu_j}$ and set
$Q_{\Lambda}(s):=\max_{j\in \Lambda} Q_j(s)$. In  \cite[Theorem~3.3]{CGK} we obtained the following extension of \eq{minamiest}:
\beq
\E \left\{ \pa{\tr  P^{(\Lambda)}_{\bom}(I)} \pa{\tr  P^{(\Lambda)}_{\bom}(I)-1} \right\}
  \le
\pa{ Q_\Lambda\pa{\abs{I}} \abs{\Lambda}}^2 .\label{minami2}
\eeq
(In \cite{CGK} the proof of \eq{minami2} is given for single-site probability measures with compact support, but \eq{minami2} follows for arbitrary single-site probability measures by Lemma~\ref{lemaprox}.  Note that the proof  is valid  for the generalized Anderson model.)

 In this article we  generalize 
Minami's estimate and its extensions to $n$ eigenvalues, allowing for
 $n$ arbitrary intervals and   arbitrary single-site probability measures with no atoms.          We also give
applications of \eq{minami2}, deriving  new results about the multiplicity of eigenvalues and Mott's
formula for the ac-conductivity when the single site probability distribution is H\"older
continuous.

The paper is organized as follows. In Section~\ref{sectresults} we state our main
results, namely our generalized 
eigenvalue-counting estimates. In Section~\ref{sectapplic}, we consider the Anderson
model with a H\"older continuous probability distribution, for which  we extend previous
results on the multiplicity of the spectrum and Mott's formula for energies in the region
of Anderson localization.  In Section~\ref{secproofs} we prove the results stated in
Section~\ref{sectresults}.  In Appendix~\ref{appW}  we provide   proofs for the fundamental spectral
averaging estimate \eq{sa}.  In Appendix~\ref{secapproxlem} we prove an approximation lemma to go from probability measures with compact support to arbitrary probability measures.

\section{Eigenvalue counting inequalities} \label{sectresults}
In this section we state our main results. The proofs will be given in Section~\ref{secproofs}.

Spectral averaging is the basic ingredient for proving  eigenvalue-counting inequalities for the generalized Anderson model.  Consider the random self-adjoint operator
\begin{equation}
H_{\omega} = H_0 + \omega \Pi_\vphi \quad \text{on}\quad \H,\label{defH10}
\end{equation}
where $H_0$ is a  self-adjoint operator on the Hilbert space $\H$, $\vphi\in \H$ with
$\norm{\vphi}=1$, and $\omega$ is a random
variable with a non-degenerate probability distribution $\mu$.   By
$\Pi_\vphi $ we denote the  orthogonal projection onto $\C \vphi$,  the one-dimensional
subspace spanned by $\vphi$.  Let $P_\omega(J)=\chi_J(H_\omega) $ for a Borel set $J
\subset \R$. There is a  fundamental spectral averaging estimate: for  all bounded
intervals $I\subset \R$ we have
\begin{equation}\label{sa}
\E_{\omega}\set{ \langle \vphi, P_\omega (I) \vphi\rangle}:= \int \d \mu(\omega)\,
\langle \vphi, P_\omega (I) \vphi\rangle \le Q_\mu\pa{\abs{I}}.
\end{equation}
  In full generality, i.e.,  $\mu$ arbitrary with $
Q_\mu(s)=8 S_\mu(s)$,  this is a recent result of Combes, Hislop and Klopp
\cite[Eq.~(3.16)]{CHK2}.  (We present a proof in Appendix~\ref{appW} for completeness.)
  If  $\mu$ has a bounded density
$\rho$, \eq{sa} was known to hold with  $ Q_\mu(s)= \norm{\rho}_\infty s$  (e.g,
\cite{W,FS,CKM,CHK,Ki};    a simple proof is given  in Appendix~\ref{appW}). If $\mu$ is
H\"older continuous,  i.e., $S_\mu(s) \le C s^\alpha$ with  $\alpha \in ]0,1[$, \eq{sa}
was known with  $ Q_\mu(s)= C ( 1- \alpha)^{-1} s^\alpha  $ \cite[Theorem~6.2]{CKM}.
 We will thus always assume that $Q_\mu$ is as in \eq{defQ}, although all we will  require of $Q_\mu$ is the validity of \eq{sa}.  (The estimate \eq{sa} is useful when  the measure $\mu$ has no atoms, i.e., $\lim_{s
\downarrow 0} Q_\mu(s)=0$, which is  always assumed in his paper.) 

Now let $H_\bom$ be the generalized Anderson model.
Note that we can rewrite the finite volume operator  given in \eq{defHfinite} as 
\begin{equation}
H_{\bom,\Lambda } = H_{0,\Lambda} + \sum_{j\in \Lambda}\omega_j  \Pi_{j} \quad
\text{on}\quad \ell^2(\Lambda), \quad \text{with $\Pi_{j}=\Pi_{\delta_j}$}. \label{defHfinite2}
\end{equation}

  The first  eigenvalue-counting inequality for $H_\bom$ is  the Wegner estimate \cite{W} , which measures the probability that $H_{\bom,\Lambda}$ has an
eigenvalue in an interval $I$:
\begin{equation}\label{wegner}
\P\set{\tr P^{(\Lambda)}_{H_\bom}(I)\ge 1}\le \E  \set{\tr P^{(\Lambda)}_{H_\bom}(I)}
\le Q_{\Lambda}\pa{\abs{I}} |\Lambda|.
\end{equation}
The Wegner estimate is an immediate consequence of \eq{sa}:
\begin{equation}
\E  \set{\tr P^{(\Lambda)}_{H_\bom}(I)}= \sum_{j \in \Lambda} \E_{\bom^\perp_j}
\set{\E_{\omega_j} \set{ \langle \delta_j,  P^{(\Lambda)}_{H_\bom}(I) \delta_j\rangle}}
\le Q_{\Lambda}\pa{\abs{I}} |\Lambda|.
\end{equation}

The second  eigenvalue-counting inequality is the Minami estimate \eq{minami2}. It is generalized to two intervals in the following theorem.

\begin{theorem}\label{thmM}   Let $H_\bom$ be the generalized Anderson model, and
fix a finite  volume $\Lambda\subset\Z^d$.  For any two
bounded intervals  $I_1, I_2$  we have
\begin{align}\label{minami}
&\E \left \{\pa{\tr P^{(\Lambda)}_{\bom}(I_1)}\pa{ \tr P^{(\Lambda)}_{\bom}(I_2)}-\min
\set{\tr P^{(\Lambda)}_{\bom}(I_1) ,\tr P^{(\Lambda)}_{\bom}(I_2)}\right \}\\
& \hspace{2.7in}\le 2\,  Q_\Lambda\pa{\abs{I_1}}Q_\Lambda\pa{\abs{I_2}}\abs{\Lambda}^2.
\notag
\end{align}
If $I_1\subset I_2$, we have
\beq\label{minamisub}
\E \left \{\pa{\tr P^{(\Lambda)}_{\bom}(I_1)}\pa{ \tr P^{(\Lambda)}_{\bom}(I_2)-1}
\right\}\le Q_\Lambda\pa{\abs{I_1}}Q_\Lambda\pa{\abs{I_2}}\abs{\Lambda}^2 .
\eeq
\end{theorem}

 \begin{remark}
 \begin{itemize}
\item[(i)] The estimate \eq{minami2}, proved in   \cite[Theorem~3.3]{CGK}, is a particular case of \eq{minamisub}.
\item[(ii)]  Note that
\begin{equation}
\pa{\tr  P^{(\Lambda)}_{\bom}(I_1)}\pa{ \tr  P^{(\Lambda)}_{\bom}(I_2)}-\min\set{ \tr
P^{(\Lambda)}_{\bom}(I_1) ,\tr  P^{(\Lambda)}_{\bom}(I_2)}\ge 0.
\end{equation}
\item[(iii)]  The intervals $I_1$ and $I_2$ in \eqref{minami} may be disjoint. In this
case the usual Minami's estimate \eq{minamiest} would yield the bound $\pi^2 \pa{\rho^{(\Lambda)}
_\infty}^2  \abs{I}^2\abs{\Lambda} ^2$, with an interval $I\supset I_1\cup I_2$, while under the same hypotheses the estimate 
\eq{minami} gives $2\pa{\rho^{(\Lambda)} _\infty}^2 |I_1|  |I_2|\abs{\Lambda}^2  $.
\end{itemize}
\end{remark}

We now turn to the general case of   $n$ arbitrary intervals, extending the results of
\cite{BHS,GV}.  Given $n \in \N$, we let  $\mathcal{S}_n$ denote the  group of all
permutations of $\set{1,2,\ldots,n}$, and recall that $\abs{\mathcal{S}_n}=n!$.  Given
 a finite  volume $\Lambda\subset\Z^d$ and    bounded intervals $I_1,\ldots ,I_n$  (not
necessarily distinct), we pick $\sigma_\bom= \sigma_\bom^{(\Lambda)}(I_1,\ldots ,I_n)\in
\mathcal{S}_n$
such that \beq \label{order}
 \tr P_\bom^{(\Lambda)}(I_{\sigma_\bom(1)})\le  \tr
P_\bom^{(\Lambda)}(I_{\sigma_\bom(2)}) \le \ldots \le \tr
P_\bom^{(\Lambda)}(I_{\sigma_\bom(n)}),
\eeq
in which case we have
\beq
\pa{\tr P_\bom^{(\Lambda)}(I_{\sigma_\bom(1)})} \pa{\tr
P_\bom^{(\Lambda)}(I_{\sigma_\bom(2)})-1}\cdots \pa{\tr
P_\bom^{(\Lambda)}(I_{\sigma_\bom(n)})-(n-1)}\ge 0 .\label{expectpos}
\eeq
To avoid ambiguity, we select $\sigma_\bom$ uniquely by requiring
$\sigma_\bom(i)<\sigma_\bom(j)$ if $i<j$ and $ \tr
P_\bom^{(\Lambda)}(I_{\sigma_\bom(i)})= \tr P_\bom^{(\Lambda)}(I_{\sigma_\bom(j)})$.
(Note that the product in the left hand side of \eqref{expectpos} is independent of the
choice of $\sigma_\bom \in \mathcal{S}_n$ satisfying \eqref{order}.)  We let $
\mathcal{S}_n(I_1,\cdots I_n)$ be the collection permutations $\sigma \in \mathcal{S}_n$
such that  $\sigma=\sigma_\bom$ for some $\bom$, and let  $M(I_1,\cdots I_n)$ denote the
cardinality of $ \mathcal{S}_n(I_1,\cdots I_n)$.  Note that   $1 \le M(I_1,\cdots I_n)\le
n!$, with  $M(I_1,\cdots I_n)=1$ if $I_1 \subset I_2\subset
\dots\subset I_n$.

\begin{theorem}\label{thmn} Let $H_\bom$ be the generalized Anderson model, 
fix a finite  volume $\Lambda\subset\Z^d$,  let
$n\in \N$, and consider  $n$ bounded intervals $I_1,\ldots, I_n$ (not necessarily
distinct).
Then, setting
$\sigma_\bom= \sigma_\bom^{(\Lambda)}(I_1,\ldots ,I_n)$,  we have
\begin{align}
& \E \left\{\pa{ \tr P_\bom^{(\Lambda)}(I_{\sigma_\bom(1)})} \pa{\tr
P_\bom^{(\Lambda)}(I_{\sigma_\bom(2)})-1}\cdots \pa{\tr
P_\bom^{(\Lambda)}(I_{\sigma_\bom(n)})-(n-1)} \right\}  \label{expectnbis}
 \\
&\qquad    \qquad  \qquad  \qquad \qquad  \qquad  \qquad  \qquad  \le
 M(I_1,\cdots I_n) \pa{ \prod_{k=1}^n Q^{(\Lambda)}\pa{\abs{I_k}}} \abs{\Lambda}^n. \notag
\end{align}
In the special case when $I_1 \subset I_2\subset \dots\subset I_n$,  we have
\begin{align}\label{expectnbis2}
&\E \left\{ \pa{\tr P_\bom^{(\Lambda)}(I_{1})} \pa{\tr P_\bom^{(\Lambda)}(I_{2})-1}\cdots
\pa{\tr P_\bom^{(\Lambda)}(I_{n})-(n-1)} \right\}
 \\
&\qquad  \quad \qquad  \qquad  \qquad  \qquad \qquad  \qquad  \qquad  \qquad  \le
 \pa{ \prod_{k=1}^n Q^{(\Lambda)}\pa{\abs{I_k}}} \abs{\Lambda}^n.  \notag
\end{align}
 In particular,  for any bounded interval  $I$ we have
\beq
\E \left\{ \pa{\tr P_\bom^{(\Lambda)}(I)} \pa{\tr P_\bom^{(\Lambda)}(I)-1}\cdots \pa{\tr
P_\bom^{(\Lambda)}(I)-(n-1)} \right\}
  \le
\pa{  Q^{(\Lambda)}\pa{\abs{I}}\abs{\Lambda} }^n . \label{expectn}
\eeq
\end{theorem}

As a corollary, we get probabilistic estimates on the number of eigenvalues of
$H_{\bom,\Lambda}$ in intervals.

\begin{corollary}\label{corcor}Let $H_\bom$ be the generalized Anderson model, and
fix a finite  volume $\Lambda\subset\Z^d$.
For all $n \in \N$ and $I$ a bounded interval, we have
\begin{equation}
\P  \set{\tr P_\bom^{(\Lambda)}(I)  \ge n }\le
\frac1{n!}\pa{   Q^{(\Lambda)}\pa{\abs{I}}\abs{\Lambda}}^n .\label{proban}
\end{equation}
Furthermore, for all bounded intervals  $I_1,\cdots I_n$  we get
\begin{equation}\begin{split}\label{proba12n}
& \P\set{ \tr P_\bom^{(\Lambda)}(I_{\sigma_\bom(1)})\ge 1, \tr
P_\bom^{(\Lambda)}(I_{\sigma_\bom(2)})\ge 2,\cdots, \tr
P_\bom^{(\Lambda)}(I_{\sigma_\bom(n)})\ge n}\\
& \qquad  \qquad  \qquad  \qquad   \qquad   \qquad \quad    \qquad  \le M(I_1,\cdots I_n)
  \pa{ \prod_{k=1}^n Q^{(\Lambda)}\pa{\abs{I_k}}}\abs{\Lambda}^n,
\end{split}\end{equation}
and, in the special case when $I_1 \subset I_2\subset \dots\subset I_n$, we have
\begin{equation}\label{proba12n2}
 \P\set{ \tr P_\bom^{(\Lambda)}(I_{1})\ge 1, \tr P_\bom^{(\Lambda)}(I_{2})\ge 2,\ldots,
\tr P_\bom^{(\Lambda)}(I_{n})\ge n}  \le  \pa{ \prod_{k=1}^n
Q^{(\Lambda)}\pa{\abs{I_k}}}\abs{\Lambda}^n .
\end{equation}
\end{corollary}

\begin{remark} Given bounded intervals $I_1$ and $I_2$, let $d(I_1,I_2)$ denote the
distance between the two intervals.  It follows from \eq{proba12n2} that
\begin{equation}\begin{split}\label{proba1-1}
& \P\set{ \tr P_\bom^{(\Lambda)}(I_{1})\ge 1\; \text{and} \;  \tr
P_\bom^{(\Lambda)}(I_{2})\ge 1}\\
 & \qquad  \le \pa{\min \set{Q^{(\Lambda)}\pa{\abs{I_1}},Q^{(\Lambda)}\pa{\abs{I_2}}}
Q^{(\Lambda)}\pa{d(I_1,I_2) + \abs{I_1} + \abs{I_2}}} \abs{\Lambda}^2 .
\end{split}\end{equation}
Note that \eq{proba1-1} does not generally hold if the right hand side is replaced by the more desirable  $C\, Q^{(\Lambda)}\pa{\abs{I_1}}Q^{(\Lambda)}\pa{\abs{I_2}}\abs{\Lambda}^2$, see \cite{AW}.
\end{remark}

\section{Applications to H\"older continuous distributions}\label{sectapplic}

The (standard) Anderson model is given
by   $H_\bom$  as in \eqref{defH}, with $H_0= -\Delta$, the centered discrete Laplacian,
and   $\bom=\{ \omega_j \}_{ j\in
\Z^d}$  a family of independent  identically distributed
random variables with joint probability distribution $\mu$, which we assume to have no
atoms.   Localization for the Anderson model has been well studied,
mostly for  $\mu$ with a bounded density $\rho$, cf. \cite{FS,FMSS,DLS,SW,DK,AM,A} and
many others, as well as for  probability distributions $\mu$ that are H\"older continuous
\cite{CKM,DK,H,ASFH,GK1}, i.e.,
$Q_\mu(s) \le U s^\alpha$ for $s$ small, for some constants $U$ and $\alpha \in ]0,1[$.
If the probability distribution $\mu$ has a bounded density, Minami's estimate
\eq{minamiest} was a crucial ingredient in Klein and Molchanov's proof of simplicity of
eigenvalues \cite{KM} and in  Klein, Lenoble and M\"uller derivation of a rigorous form
of Mott's formula for the ac-conductivity \cite{KLM}.  In this section we show that with
\eq{minami2} these proofs extend to the case when $\mu$ is only H\"older continuous.

\subsection{Multiplicity of the spectrum}

Let $H_\bom$ be a generalized Anderson model as in \eqref{defH}, let $\alpha \in ]0,1]$,
and assume that
the probability distributions $\mu_j$ are uniformly $\alpha$-H\"older continuous, i.e.,
there is a constant $U$ and $s_0 >0$ such that
\beq \label{Hconts}
\sup_{j \in \Z^d} Q_j(s) \le U s^\alpha  \quad \text{for all} \quad s \in [0,s_0].
\eeq
 In this case we say that $H_\bom$ is an $\alpha$-H\"older continuous generalized
Anderson model.

We  say that the generalized Anderson model $H_\bom$ exhibits Anderson localization in
some interval $I$ if, with probability one, $H_\bom$ has pure point spectrum in $I$ and
the  corresponding eigenfunctions decay exponentially.  Given $x\ge 0$, we let  $[x]$
denote the integer part of $x$.  Following Klein and Molchanov \cite{KM}, we prove the
following result.

\begin{theorem} \label{thmmult} Let $H_\bom$ be an  $\alpha$-H\"older continuous
generalized Anderson model.  Suppose $H_\bom$ exhibits Anderson localization in some
interval $I$.
 Then, with probability one,
  every eigenvalue of $H_\bom $ in $I$ has multiplicity $ \le [\alpha^{-1}]$.
  In particular, if $\alpha>\frac12$, with probability one  every eigenvalue of $H_\bom $
in $I$ is simple.
\end{theorem}

\begin{remark} For the standard Anderson model, where  $\mu_j=\mu$ for all $j \in \Z^d$,
this theorem was originally proved by Simon  \cite{Si} when $\mu$ has a bounded density
(i.e., $\alpha=1$).   Our proof is based on  a simple proof later provided by Klein and
Molchanov \cite{KM}, based on Minami's estimate \eq{minamiest}.   For singular measures
(i.e., $\alpha <1$), the best previously known result for the standard Anderson model  is
the finite multiplicity of the eigenvalues \cite{CH,GK};  uniform boundedness of the
multiplicity was not previously known.  Thus Theorem~\ref{thmmult} improves on both
results.
\end{remark}

\begin{proof}[Proof of Theorem~\ref{thmmult}]
We proceed as in  \cite{KM}.     We call
 $\vphi \in \ell^2(\Z^d)$
$\alpha$-fast decaying if it has  $\beta$-decay, that is,
 $\abs{\vphi(x)}
\le C_\vphi \pa{1+|x|}^{-\beta}$ for some  $C_\vphi < \infty$, with
\beq \label{betadec}
\beta > \pa{ \frac 1 2 +\pa{ \alpha -\pa{[\alpha^{-1}] +1 }^{-1}     }^{-1}   }d.
\eeq
To prove the theorem, we will show that, with probability one, an $\alpha$-H\"older
continuous generalized Anderson model
$H_\bom$  cannot  have an eigenvalue with $[\alpha^{-1}] + 1$  linearly independent
$\alpha$-fast decaying eigenfunctions.

 We set $N=[\alpha^{-1}]+1$, so that $N\alpha > 1$. For a given open interval
$I$ we  pick
\beq \label{qq}
q >  \frac {Nd}{N \alpha -1}=\pa{ \alpha -\pa{[\alpha^{-1}] +1 }^{-1}     }^{-1} d.
\eeq
Given a scale $L >0$,  we let
$\Lambda_L$ denote the cube of side $L$ centered at $0$,
and cover $I$
 by $2 \left (\left[\frac {L^q} 2 |I|\right]+1\right)\le
{L^q}  |I| +2 $
intervals of length $2L^{-q} $, in such a way that
any subinterval $J \subset I $
with length $|J| \le L^{-q} $ will be contained in one of these intervals.
We consider the event
 $\mathcal{B}_{L,I,q}$, which occurs if
there exists an  interval
 $J \subset I $
with  $|J| \le L^{-q} $ such that   $\tr P_\bom^{(\Lambda_L)}(J) \ge N $.  Its probability
can be estimated, using
\eq{proban} and \eq{Hconts}, by
\begin{equation}
\P\{\mathcal{B}_{L,I,q}\}\le  \tfrac 1 {N!} ( L^q |I|+2) \pa{U \pa{2L^{-q}}^\alpha L^d}^N
\le  (|I| +1)\tfrac {(2^\alpha U)^N}{N!}  L^{-(N \alpha-1) q+ Nd}.
\end{equation}
In view of \eq{qq},  taking scales $L_k=2^k$, it follows  from the Borel-Cantelli Lemma
that, with probability one,   the event   $\mathcal{B}_{L_k,I,q}$
eventually does not occur.

Now, suppose that for some $\bom$  there exists  $E\in I$ which  is an eigenvalue of
$H_\bom$ with   $N$ linearly independent
$\alpha$-fast decaying eigenfunctions, so they all have $\beta$-decay for some $\beta$ as
in \eq{betadec}.  It follows,  as in \cite[Lemma~1]{KM},  that for   $L$ large enough the
finite volume operator  $H_{\bom, \Lambda_L}$  has at least $N $ eigenvalues in the
interval $J_{E,L}=[E-\eps_L,E+\eps_L]$, where $\eps_L= C L^{-\beta +\frac d 2}$ for an
appropriate  constant $C$ independent of $L$.  In view of \eq{betadec}, we can pick $q$,
satisfying \eq{qq}, such that
$\beta - \frac d 2 > q$, and hence  $\eps_L < L^{-q}$ for all large $L$.  But  with
probability one this is impossible since  the event   $\mathcal{B}_{L_k,I,q}$
 does not occur for large $L_k$.
\end{proof}

\subsection{Generalized Mott's formula}

Let $\alpha \in ]0,1[$, and  consider the Anderson model
$H_\bom$ with a single-site probability distribution $\mu$ of compact support
and  uniformly  $\alpha$-H\"older continuous:
\beq \label{Hconts2}
 Q_\mu(s) \le U s^\alpha  \quad \text{for all} \quad s \in [0,s_0],
\eeq
where $U$ and $s_0 >0$ are constants.  The fractional moment method can be applied to
such measures, leading to exponential decay of the expectation of some fractional power
of the Green's function \cite{H,ASFH}. We may then define the region of complete
localization  $\Xi^{\mathrm{CL}}$, introduced in  \cite{GKduke,GK}, as in \cite[Definition~2.1]{KLM}. However,  \cite[Eqs.
(4.1), (4.3)  and (4.4)]{KLM} have not been derived from the fractional moment method for
$\mu$ with compact support  satisfying only  the condition \eq{Hconts2}.
(\cite[Appendix~A]{H,ASFH} assumes that  $\mu$ thas a bounded density in the  derivation
of  such estimates.)  But in this  region of complete localization we can always perform
a multiscale analysis as in \cite{GK1} with only hypothesis \eq{Hconts2}, and get the
estimates  \cite[Eqs. (4.1), (4.3)]{KLM} with sub-exponential decay  \cite{GK1,GK}, and
hence conclude that  \cite[Assumption~3.1 and Eq.~(4.4)]{KLM} are satisfied.  Thus,
  given  a Fermi energy  $E_{F} \in \Xi^{\mathrm{CL}}$, the analysis in \cite{KLM}
applies and we may define the \emph{average
  in phase conductivity} $ \overline{\sigma}_{E_{F}}^{\mathrm{in}}(\nu)$ as in
\cite[Eq.~(2.17)]{KLM}.  We have the following extension of \cite[Theorem~2.3]{KLM}.
Note that we get  $ \overline{\sigma}_{E_{F}}^{\mathrm{in}}(\nu)
  \le C  \nu^{2\alpha} \left(\log\tfrac{1}{\nu}\right)^{d+2} $  for small $\nu$,
consistent with
  $ C  \nu^{2} \left(\log\tfrac{1}{\nu}\right)^{d+2} $ for $\alpha=1$ as in \cite{KLM}.

\begin{theorem}
  \label{alphamott} Given  $\alpha \in ]0,1[$, let
 $H_\bom$ be an   Anderson model
with a single-site probability distribution $\mu$ of compact support
and  uniformly  $\alpha$-H\"older continuous   as in
\eq{Hconts2}.  Consider a Fermi energy in its
  region of complete localization: $E_{F} \in \Xi^{\mathrm{CL}}$. Then
  \begin{equation}
    \label{leadingbound8}
    \limsup_{\nu \downarrow 0} \frac {
      \overline{\sigma}_{E_{F}}^{\mathrm{in}}(\nu)}
    { \nu^{2\alpha} \left(\log\tfrac{1}{\nu}\right)^{d+2}} \le B \, {\ell}_{E_{F}}^{d+2} ,
  \end{equation}
  where $\ell_{E_{F}}$ is given in \cite[Eq.~(2.3)]{KLM},  and the constant $B$ depends
only on $d$, $U$ and $\alpha$.
\end{theorem}

\begin{proof} The proof of \cite[Theorem~2.3]{KLM} applies, with modifications due to the
use of \eq{wegner} and \eq{minami2} with $Q(s)=Q_\mu(s)$ as in \eq{Hconts2}. The
modifications are as follows (we use the notation of  \cite{KLM}):

\begin{enumerate}
\item  We systematically use    \eq{sa} instead of \cite[Eq.~(4.5)]{KLM}. In particular,
\cite[Eq.~(4.10)]{KLM} becomes
 \begin{equation}
    \la\la Y_{E_{F}}, \chi_{B}(\mathcal{H}_{L}) Y_{E_{F}} \ra\ra \le
    W_{\beta}\, Q( \abs{B})^{\beta} \quad \text{for all Borel sets $B\subset\R$}.
  \end{equation}
To derive this estimate, we use the sub-exponential decay  of the Fermi projection given
in  \cite[Theorem~3]{GK}, i.e.,  we use  \cite[Eq.~(4.1)]{KLM} but with sub-exponential
decay.
This can be done because we only use summability of this decay.
As a consequence,  \cite[Eq.~(4.6)]{KLM} becomes
 \begin{equation}
    \Psi_{E_{F}}(B_{+}\times B_{-})
    \le W_{\beta}\pa{\min\set{ Q(|B_{+}|), Q(|B_{-}|}}^{\beta}.
      \end{equation}

\item  Using \eq{minami2},   \cite[Eq.~(4.51)]{KLM} becomes
  \begin{equation}\label{finitevolumeMott}
    \E \set {\langle\delta_{0}, F_{-,L} X_{1} F_{+,L}X_{1}F_{-,L}
    \delta_{0}\rangle }  \le   \tfrac{ 1} 4
    Q(|J|)^{2} L^{d+2} .
  \end{equation}

\item  In  \cite[Lemma~4.6]{KLM}, we cannot use the estimate   \cite[Eq.~(4.28)]{KLM}.
But proceeding as in the proof of  \cite[Eq.~(4.30)]{KLM}, we can replace it by
\begin{equation}\label{Fdecay}
    \E \left \{ \left\lvert\langle\delta_{x},F_{\pm}
        \delta_{y}\rangle\right\rvert^{p}\right\}
    \le C_ {I} \hnorm{f_\pm}_2\e^{- \frac {1} {\ell} \abs{x-y}}  \quad \text{for all $p
\in
      [1,\infty[$  and  $ x,y \in \Z^{d}$}.
  \end{equation}
As a consequence, the right hand side of  \cite[Eq.~(4.27)]{KLM} becomes
\beq
 C \pa{  \pa{ \hnorm{f_{+}}_2\hnorm{f_{-}}_3\hnorm{f_{-}}_4}^{\frac 1 3}+    \pa{
\hnorm{f_{-}}_3^2\hnorm{f_{+}}_4}^{\frac 1 3}    } L^{\frac 4 3 d}
      \; \e^{-\frac {1} {12 \ell} L}.
\eeq

\item  Using \eq{minami2} instead of \cite[Eq.~(4.47)]{KLM}, and taking into account  the
above modifications, \cite[Eq.~(4.61)]{KLM} becomes
\begin{align}
  \Psi_{E_{F}}(I_{+}\times I_{-}) & \le
 \tfrac 1 4 Q(2 \nu)^2 L^{d+2}+
  C^{\prime} \nu^{{-15}}  L^{\frac 4 3 d} \;
  \e^{-\frac {1} {12\ell} L} +4W_{\frac 1 2} Q(\nu^{4})^{\frac 1 2}\\
  & \le 4^{\alpha-1} U^{2} \nu^{2\alpha} L^{d+2} +   C^{\prime} \nu^{{-15}}  L^{\frac 4 3
d} \;
  \e^{-\frac {1} {12\ell} L} + 4U^{\frac 1 2}W_{\frac 1 2}  \nu^{2\alpha},\notag
\end{align}
where we used \eq{Hconts2}.

\item  As in \cite[Eq.~(4.62)]{KLM}, we choose $L=  A \ell \log \frac 1 \nu$, where $A$
is some suitable constant, depending on $d$, $\alpha$ and $U$, such that, similarly to
\cite[Eq.~(4.63)]{KLM}, we get
\begin{align}\label{finalest}
  \Psi_{E_{F}}(I_{+}\times I_{-})  \le B
 \ell^{d +2} \nu^{2\alpha} \left(\log
    \tfrac{1}{\nu}\right)^{d+2}+
  C^{\prime\prime}  \nu^{2\alpha},
\end{align}
where $B$ and $ C^{\prime\prime} $ are constants, with $B$ depending only on $d$,
$\alpha$ and $U$, from which \eqref{leadingbound8} follows.

\end{enumerate}

\end{proof}

%%%%%%%%%%%%%%%%%%%%%%%%%%%%%
%%%%%%%%%%%%%%%%%%%%%%%%%%%%%%%%%%%%%%%%%
%%%%%%%%%%%%%%%%%%%%%%%%%%%%%%%%%%%%%%%%%
%%%%%%%%%%%%%%%%%%%%%%%%%%%%%%%%%%%%%%%%%
%%%%%%%%%%%%%%%%%%%%%%%%%%%%%%%%%%%%%%%%%

\section{Proofs of  eigenvalue counting inequalities}\label{secproofs}

In this section we prove Theorems~\ref{thmM} and \ref{thmn} and Corollary~\ref{corcor}. Since we always have  $\tr P_\bom(I) \le \abs{\Lambda}$, it follows from  Lemma~\ref{lemaprox}
that it suffices to prove
 the theorems   when  all the  probability measures $\mu_j$ have compact support, which   is assumed in the proofs of Theorems~\ref{thmM} and \ref{thmn}.

Our proofs are based on  the fundamental spectral averaging estimate \eq{sa} and \cite[Lemma~3.2]{CGK}, which we now state.

\begin{lemma}[\cite{CGK}]\label{lemkey} Consider the self-adjoint operator  $
H_{s} = H_0 + s \Pi_\vphi $ on  the Hilbert space $ \H$,
where $H_0$ is a  self-adjoint operator on  $\H$, $\vphi\in \H$ with $\norm{\vphi}=1$,
and $s\in \R$.  Let $P_s(J)= \chi_J(H_s)$ for an interval $J$, and suppose
$\tr P_0(]-\infty,c])<\infty$ for all $c \in \R$.  Then, given $a,b \in \R$ with $a <b$,
we have
\beq
\tr P_s(]a,b])\le 1 + \tr P_t(]a,b]) \quad \text{for all $0\le s \le t$}. \label{decomp}
\eeq
\end{lemma}

Let $\Lambda \subset \Z^d$ finite .  Given $\bom \in \R^{\Z^d}$, we set
 $P_\bom^{(\Lambda)}(I)=\chi_I(H_{\bom,\Lambda})$.  Given $j \in \Lambda$,  we
write $\bom =(\bom_j^\perp,\omega_j)$ and   $P^{(\Lambda)}_{\omega_j=s}(I)=P^{(\Lambda)}_{(\omega_j^\perp,s)}(I)$
when we want to make explicit the value of $\omega_j$.
We also write $P^{(\Lambda)}_{\omega_j\to s}(I)$ to denote that $\omega_j$ was replaced by $s$.

Since we assumed that the measures $\mu_j$ have no atoms, it follows from  \eq{wegner}
that $\E_\bom\set{ \tr P_\bom^{(\Lambda)}(\{c\})}=0$ for any $c \in \R$. Thus it does not
matter if the intervals are open or closed at the endpoints, so  in the proofs we may
take all intervals to be of the form $]a,b]$,  which allows the use of Lemma~\ref{lemkey}.

Theorem~\ref{thmM} is a particular case of Theorem~\ref{thmn},
 but in order to illustrate the simplicity of our approach we first give a proof of
Theorem ~\ref{thmM} and then prove the general case.

\begin{proof}[Proof of Theorem~\ref{thmM}] Fix a finite volume $\Lambda\subset \Z^d$ and let $I_1$, $I_2$
be bounded intervals.  Using Lemma~\ref{lemkey}, for $\tau_j \ge \omega_j$ we always  have
\begin{align}
\pa{\tr P\up{\Lambda}_\bom(I_1)} (\tr P\up{\Lambda}_\bom(I_2)-1)
& =
\sum_{j\in\Lambda} \set{ \scal{\delta_j, P\up{\Lambda}_\bom(I_1) \delta_j}\pa{\tr P\up{\Lambda}_\bom(I_2)-1}  } \\
& \le
\sum_{j\in\Lambda} \set{ \scal{\delta_j, P\up{\Lambda}_{(\bom_j^\perp,\omega_j)}(I_1) \delta_j}
\pa{\tr P\up{\Lambda}_{(\bom_j^\perp,\tau_j)}(I_2)}}. \notag
\end{align}
We now take  $\tau_j \ge \max \supp \mu_j$ for all $j\in \Lambda$, and average over the
random variables $\bom=\{ \omega_j \}_{ j\in
\Z^d}$, where each $\omega_j$ has the probability distribution $\mu_j$.   Using
\eqref{sa}, we get
\begin{align}
&\E_\bom \set{\pa{\tr P\up{\Lambda}_\bom(I_1)}(\tr P\up{\Lambda}_\bom(I_2)-1) }\\
& \qquad \qquad \le
\sum_{j\in\Lambda} \E_{\bom_j^\perp} \set{  \pa{\tr P\up{\Lambda}_{(\bom_j^\perp,\tau_j)}(I_2)
}\pa{\E_{\omega_j} \set{\scal{\delta_j, P\up{\Lambda}_{(\bom_j^\perp,\omega_j)}(I_1) \delta_j} } }
}\notag \\
& \qquad \qquad \le
Q_\Lambda\pa{\abs{I_1}} \, \sum_{j\in\Lambda} \E_{\bom_j^\perp} \set{\tr
P\up{\Lambda}_{(\bom_j^\perp,\tau_j)}(I_2)}.\notag
\end{align}
This holds for all  $\tau_j \ge \max \supp \mu_j$, $j\in \Lambda$, so we now take
$\tau_j= \max \supp \mu_j + \tilde{\omega}_j$, where $\tilde{\bom}=\set{
\tilde{\omega}_j}_{j\in\Z^d}$ and  $\bom=\set{{\omega}_j}_{j\in\Z^d}$ are two independent,
identically distributed collections of random variables, and average over these random
variables. We get
\begin{align}\notag
&\E_\bom \set{\pa{\tr P\up{\Lambda}_\bom(I_1)}(\tr P\up{\Lambda}_\bom(I_2)-1) }=\E_{\tilde{\bom}}\!\set{  \E_\bom
\set{\pa{\tr P\up{\Lambda}_\bom(I_1)}(\tr P\up{\Lambda}_\bom(I_2)-1) }\!}\\
&\qquad \quad  \le Q_\Lambda\pa{\abs{I_1}} \sum_{j\in\Lambda}
\E_{(\bom_j^\perp,\tilde{\omega}_j)} (\tr P\up{\Lambda}_{(\bom_j^\perp,\tau_j)}(I_2)) \le
Q_\Lambda\pa{\abs{I_1}}Q_\Lambda\pa{\abs{I_2}} |\Lambda|^2, \label{pre-minami}
\end{align}
where we used  the  Wegner estimate \eqref{wegner}.

The estimates \eqref{minamisub} and \eqref{minami2}  follow immediately from
\eq{pre-minami}. To get \eqref{minami}, we use \eqref{pre-minami} and the obvious estimate
\begin{align}
&\pa{\tr  P\up{\Lambda}_\bom(I_1)}\pa{ \tr  P\up{\Lambda}_\bom(I_2)}-\min \set{\tr P\up{\Lambda}_\bom(I_1) ,\tr
P\up{\Lambda}_\bom(I_2)}\\
& \qquad  \qquad \le \pa{\tr P\up{\Lambda}_\bom(I_1)}\pa{\tr P\up{\Lambda}_\bom(I_2)-1} +\pa{ \tr
P\up{\Lambda}_\bom(I_2)}\pa{\tr P\up{\Lambda}_\bom(I_1)-1}. \notag
\end{align}
\end{proof}

We now turn to the general case.

\begin{proof}[Proof of Theorem~\ref{thmn}] Fix a finite volume $\Lambda\subset \Z^d$.
We  first prove \eqref{expectn},  a particular case of  \eqref{expectnbis}, since  it has a simpler
proof.   We fix the bounded interval $I$ and proceed by
induction on $n$. The case $n=1$,  is just Wegner's inequality \eqref{wegner}.  Let us
assume that \eqref{expectn} holds for $n$, for all possible probability distributions
$\mu_j$ with compact support.  Then, given $j\in\Lambda$ and  $\tau_j \ge \max \supp
\mu_j$,  we have, using  \eqref{decomp}, that  for all $k=1,2,\ldots,n$,
\begin{equation}
\tr P\up{\Lambda}_\bom(I) - k \le  1 + \tr P\up{\Lambda}_{(\bom_j^\perp,\tau_j)}(I) -k = \tr
P\up{\Lambda}_{(\bom_j^\perp,\tau_j)}(I) -(k-1).
\end{equation}
Note that $\pa{ \tr P\up{\Lambda}_{(\bom_j^\perp,\tau_j)}(I)} (\tr P\up{\Lambda}_{(\bom_j^\perp,\tau_j)}(I)-1)
\cdots (\tr P\up{\Lambda}_{(\bom_j^\perp,\tau_j)}(I)-(n-1))\ge 0$.  Since either $\tr P\up{\Lambda}_\bom(I) (\tr
P\up{\Lambda}_\bom(I)-1) \cdots (\tr P\up{\Lambda}_\bom(I)-n)=0$ or  $\tr P\up{\Lambda}_\bom(I) - k >0$ for $k=0,1,\ldots,n$,
 it follows that we always have
\begin{align} \label{prodtr}
&\pa{\tr P\up{\Lambda}_\bom(I)} (\tr P\up{\Lambda}_\bom(I)-1) \cdots (\tr P\up{\Lambda}_\bom(I)-n)  \\
& \le
\sum_{j\in\Lambda}
\langle \delta_j, P\up{\Lambda}_\bom(I) \delta_j \rangle ( \tr P\up{\Lambda}_{(\bom_j^\perp,\tau_j)}(I)) (\tr
P\up{\Lambda}_{(\bom_j^\perp,\tau_j)}(I)-1) \!\cdots \! (\tr P\up{\Lambda}_{(\bom_j^\perp,\tau_j)}(I)-(n-1)). \notag
\end{align}
Using \eq{sa}, we have
\begin{align}\notag
&\E \set{\!\langle \delta_j, P\up{\Lambda}_\bom(I) \delta_j \rangle ( \tr P\up{\Lambda}_{(\bom_j^\perp,\tau_j)}(I))
(\tr P\up{\Lambda}_{(\bom_j^\perp,\tau_j)}(I)-1)\! \cdots\! (\tr P\up{\Lambda}_{(\bom_j^\perp,\tau_j)}(I)-(n-1))\!}\\
\label{prodtr2}
& \; = \E_{\bom_j^\perp} \left\{\E_{\omega_j}\set{\langle \delta_j, P\up{\Lambda}_\bom(I) \delta_j
\rangle}  (\tr P\up{\Lambda}_{(\bom_j^\perp,\tau_j)}(I))    (\tr P\up{\Lambda}_{(\bom_j^\perp,\tau_j)}(I)-1)
\times \right.\\& \hspace{3in} \cdots
\left.(\tr P\up{\Lambda}_{(\bom_j^\perp,\tau_j)}(I)-(n-1))\right\} \notag\\
& \; \le
Q_\Lambda(\abs{I}) \,   \E_{\bom_j^\perp}\! \left\{  (\tr P\up{\Lambda}_{(\bom_j^\perp,\tau_j)}(I))
(\tr P\up{\Lambda}_{(\bom_j^\perp,\tau_j)}(I)-1) \! \cdots \!
(\tr P\up{\Lambda}_{(\bom_j^\perp,\tau_j)}(I)-(n-1))\!\right\}.
\notag
\end{align}
 We now take $\btau= \set{\tau_{j}= a_j + \tilde{\omega}_j}_{j \in \Lambda}$,
where  $\tilde{\bom}= \set{\tilde{\omega}_j}_{j \in \Lambda}$
are independent  random variables, independent of $\bom$, such that
$\tilde{\omega}_j$   has $\mu_{j}$ for probability distribution,  and  $ a_j=  \max \supp
\mu_j $. Using \eq{prodtr} and \eq{prodtr2}, plus the induction hypothesis, we get
\begin{align}
&\E_\bom \set{ \tr P\up{\Lambda}_\bom(I) (\tr P\up{\Lambda}_\bom(I)-1) \cdots (\tr P\up{\Lambda}_\bom(I)-n)}\notag \\
& \qquad = \E_{(\bom,\btau)} \set{ \tr P\up{\Lambda}_\bom(I) (\tr P\up{\Lambda}_\bom(I)-1) \cdots (\tr
P\up{\Lambda}_\bom(I)-n)}\\
& \qquad \le Q_\Lambda(\abs{I})
 \sum_{j\in\Lambda}   \E_{(\bom_j^\perp, \tau_j)} \left\{ (\tr
P\up{\Lambda}_{(\bom_j^\perp,\tau_j)}(I))    (\tr P\up{\Lambda}_{(\bom_j^\perp,\tau_j)}(I)-1) \times
\right.\notag\\
 &\hspace{3in} \left. \cdots
(\tr P\up{\Lambda}_{(\bom_j^\perp,\tau_j)}(I)-(n-1))\right\} \notag\\
&\qquad \le
Q_\Lambda(\abs{I})\sum_j (Q_\Lambda(\abs{I})  |\Lambda|)^{n}
=
\pa{Q_\Lambda(\abs{I}) |\Lambda|}^{n+1} .\notag
\end{align}

We now turn to the proof of  \eqref{expectnbis}.  The case $n=1$ is just \eq{wegner}, and
$n=2 $ is \eq{minami}, so we assume $n\ge 3$.    Let $I_1,I_2,\ldots,I_n$ be bounded
intervals. For a fixed $\bom$, we have \eq{order}
and \eqref{expectpos}.  Let us suppose
\beq
\Xi_\bom:=\pa{ \tr P\up{\Lambda}_\bom(I_{\sigma_\bom(1)})} \pa{\tr
P\up{\Lambda}_\bom(I_{\sigma_\bom(2)})-1}\cdots \pa{\tr
P\up{\Lambda}_\bom(I_{\sigma_\bom(n)})-(n-1)} > 0,
 \eeq
 and note that in this case  we    must  have
\beq
\tr P\up{\Lambda}_\bom(I_{\sigma_\bom(k)})-k+1\ge 1, \quad \text{i.e.}, \quad \tr
P\up{\Lambda}_\bom(I_{\sigma_\bom(k)})\ge k ,  \quad
\text{ for all $k=1,\cdots,n$}.
\eeq
Then, using Lemma~\ref{lemkey} repeatedly, we get
\begin{align}\notag
\Xi_\bom& \le \sum_{j_1 \in \Lambda}\set{ \scal{\delta_{j_1}, P\up{\Lambda}_\bom
(I_{\sigma_\bom(1)})\delta_{j_1}} \pa{\tr P\up{\Lambda}_{\bom^{(j_1)}}(I_{\sigma_\bom(2)})}\cdots
\pa{\tr P\up{\Lambda}_{\bom^{(j_1)}}(I_{\sigma_\bom(n)})-n} }
\\
& \le \ldots\ldots\ldots\ldots  \label{bigmess}
 \\ \notag
& \le \sum_{j_1, j_2,\ldots, j_{n-1} \in \Lambda}\left\{ \scal{\delta_{j_1},
P\up{\Lambda}_{\bom}(I_{\sigma_\bom(1)}) \delta_{j_1}}
\scal{\delta_{j_2}, P\up{\Lambda}_{\bom^{(j_1)}}(I_{\sigma_\bom(2)}) \delta_{j_2}} \cdots \times
\right. \\
 &\qquad \quad \quad \left.\scal{\delta_{j_{n-1}}, P\up{\Lambda}_{\bom^{(j_1, j_2,\ldots,
j_{n-2})}}(I_{\sigma_\bom(n-1)}) \delta_{j_{n-1}}}\pa{\tr P\up{\Lambda}_{\bom^{(j_1, j_2,\ldots,
j_{n-1})}}(I_{\sigma_\bom(n)})}\right\},\notag
\end{align}
where \  $\bom^{(j_1)}$ is $\bom$ with $\omega_{j_1}\to \tau^{(1)}_{j_1}$, \
$\bom^{(j_1,j_2)}$ is $\bom^{(j_1)}$ with $\omega^{(j_1)}_{j_2}\to \tau^{(2)}_{j_2}$,
\ldots, \
$\bom^{(j_1, j_2,\ldots, j_{n-1})}$ is $\bom^{(j_1, j_2,\ldots, j_{n-2})}$ with
$\omega^{(j_1, j_2,\ldots, j_{n-2})}_{j_{n-1}}\to \tau^{(n-1)}_{j_{n-1}}$.  To be able to
apply  Lemma~\ref{lemkey} we must have
\beq\label{taucond}
\omega_{j_1}\le  \tau^{(1)}_{j_1}, \omega_{(j_2)}^{(j_1)}\le  \tau^{(2)}_{j_2},\ldots,
\omega^{(j_1, j_2,\ldots, j_{n-2})}_{j_{n-1}}\le \tau^{(n-1)}_{j_{n-1}},  \quad   j_k \in
\Lambda, \ k=1,2,\ldots,n-1.
\eeq
 We then take
$$\btau= \set{\tau^{(k)}_{j_k}= a^{(k)}_{j_k} + \omega^{(k)}_{j_k}; \  j_k \in \Lambda, \
k=1,2,\ldots,n-1},$$
where  $\widehat{\bom}= \set{\omega^{(k)}_{j_k}; \  j_k \in \Lambda, \ k=1,2,\ldots,n-1}$
are independent  random variables, independent of $\bom$, such that
$\omega^{(k)}_{j_k}$   has $\mu_{j_k}$ for probability distribution,  and the  real
numbers  $ a^{(k)}_{j_k} $  are chosen such that \eq{taucond} holds
$\P_{(\bom,\widehat{\bom})}$-almost surely.

Since the last expression in \eq{bigmess} is obviously nonnegative, it follows that
 \eq{bigmess} holds also when $\Xi_\bom=0$, an hence it holds
$\P_{(\bom,\widehat{\bom})}$-almost surely.

Given $\sigma \in \mathcal{S}_n$, let
\begin{align}
&\Xi_{\bom,\btau,\sigma}:=  \sum_{j_1, j_2,\ldots, j_{n-1} \in \Lambda}\left\{
\scal{\delta_{j_1}, P\up{\Lambda}_{\bom}(I_{\sigma(1)}) \delta_{j_1}}
\scal{\delta_{j_2}, P\up{\Lambda}_{\bom^{(j_1)}}(I_{\sigma(2)}) \delta_{j_2}} \cdots \times \right. \\
 & \qquad  \qquad \qquad  \left.\scal{\delta_{j_{n-1}}, P\up{\Lambda}_{\bom^{(j_1, j_2,\ldots,
j_{n-2})}}(I_{\sigma(n-1)}) \delta_{j_{n-1}}}\pa{\tr P\up{\Lambda}_{\bom^{(j_1, j_2,\ldots,
j_{n-1})}}(I_{\sigma(n)})}\right\} \notag .
\end{align}
It follows that $\P_{(\bom,\widehat{\bom})}$-almost surely  we have
\begin{align}\label{oversigma}
\Xi_\bom \le \sum_{\sigma \in  \mathcal{S}_n(I_1,\cdots I_n)} \Xi_{\bom,\btau,\sigma},
\end{align}
and hence
\begin{align}
\E_\bom \set{\Xi_\bom}= \E_{(\bom,\widehat{\bom})} \set{\Xi_\bom}\le \sum_{\sigma \in
\mathcal{S}_n(I_1,\cdots I_n)} \E_{(\bom,\widehat{\bom})}\set{ \Xi_{\bom,\btau,\sigma}}.
\end{align}
By performing the integrations in the right order, using \eqref{sa} $n-1$ times, and then
using the  Wegner estimate \eqref{wegner}, we get
\beq
 \E_{(\bom,\widehat{\bom})} \set{  \Xi_{\bom,\btau,\sigma}}\le  \pa{ \prod_{k=1}^n
Q^{(\Lambda)}\pa{\abs{I_k}}} \abs{\Lambda}^n .
 \eeq
Since $M(I_1,\cdots I_n)=\abs{ \mathcal{S}_n(I_1,\cdots I_n)}$, the estimate
\eqref{expectnbis} follows.
\end{proof}

%%%%%%%%%%%%%%%%%%%%%%%%%%%%%%%%%%%%%%%%%

\begin{proof}[Proof of Corollary~\ref{corcor}]
The estimate \eqref{proban} follows from  \eq{expectn} and the inequality
\begin{align}\notag
&\P\set{ \tr P\up{\Lambda}_\bom(I) \ge n}\\
&\qquad  \le
\P\set {\pa{\tr P\up{\Lambda}_\bom(I)} (\tr P\up{\Lambda}_\bom(I)-1) \cdots (\tr P\up{\Lambda}_\bom(I)-(n-1)) \ge n!} \\
&\qquad  \le
\frac1{n!} \E\set{\pa{ \tr  P\up{\Lambda}_\bom (I) |}(\tr P\up{\Lambda}_\bom (I)-1) \cdots (\tr P\up{\Lambda}_\bom (I)-(n-1)}\notag .
\end{align}

To obtain \eq{proba12n}, we use \eq{expectnbis}  with
\begin{align}
 & \P\set{ \tr P_\bom^{(\Lambda)}(I_{\sigma_\bom(1)})\ge 1, \tr
P_\bom^{(\Lambda)}(I_{\sigma_\bom(2)})\ge 2,\cdots, \tr
P_\bom^{(\Lambda)}(I_{\sigma_\bom(n)})\ge n}\\
 & \quad\le  \E \left\{ \pa{\tr P_\bom^{(\Lambda)}(I_{\sigma_\bom(1)}) }\pa{\tr
P_\bom^{(\Lambda)}(I_{\sigma_\bom(2)})-1}\cdots \pa{\tr
P_\bom^{(\Lambda)}(I_{\sigma_\bom(n)})-(n-1)} \right\} . \notag
\end{align}
Similarly, \eq{proba12n2} follows from \eq{expectnbis2}.
\end{proof}

%%%%%%%%%%%%%%%%%%%%%%%%%%%%%%%%%%%%%%%%%
%%%%%%%%%%%%%%%%%%%%%%%%%%%%%%%%%%%%%%%%%
%%%%%%%%%%%%%%%%%%%%%%%%%%%%%%%%%%%%%%%%%
%%%%%%%%%%%%%%%%%%%%%%%%%%%%%%%%%%%%%%%%%
%%%%%%%%%%%%%%%%%%%%%%%%%%%%%%%%%%%%%%%%%
%%%%%%%%%%%%%%%%%%%%%%%%%%%%%%%%%%%%%%%%%
%%%%%%%%%%%%%%%%%%%%%%%%%%%%%%%%%%%%%%%%%
%%%%%%%%%%%%%%%%%%%%%%%%%%%%%%%%%%%%%%%%%
%%%%%%%%%%%%%%%%%%%%%%%%%%%%%%%%%%%%%%%%%
%%%%%%%%%%%%%%%%%%%%%%%%%%%%%%%%%%%%%%%%%
\appendix

\section{The fundamental spectral averaging estimate}\label{appW}

For the reader convenience we present a proof of the fundamental spectral averaging
result \eqref{sa}.  Consider the random self-adjoint operator  on a Hilbert space $\H$ given in \eq{defH10}: $H_\omega=H_0+\omega\Pi_\vphi$,
where $H_0$ is a  self-adjoint operator on the Hilbert space $\H$, $\vphi\in \H$ with
$\norm{\vphi}=1$, $\Pi_\vphi $ is the  orthogonal projection onto   the one-dimensional
subspace spanned by $\vphi$, and $\omega$ is a random
variable with a non-degenerate probability distribution $\mu$. 
   Given $z$ with $\Im z>0$, we have,  as in \cite[Proof of
Lemma~6.1]{CKM}, that
\begin{equation}\label{rank1}
\langle \vphi, (H_\omega-z)^{-1}\vphi\rangle = \pa{\la\vphi,(H_0-z)^{-1} \vphi\ra^{-1}  +
\omega }^{-1}.
\end{equation}

\subsection{The probability distribution $\mu$ has a bounded density $\rho$}
   In this case,  we use
\begin{equation}
\int_\R \d \omega \,  \Im\langle \vphi, (H_\omega-z)^{-1}\vphi \rangle = \pi ,
\end{equation}
 a consequence of \eq{rank1} (cf. \cite[Proof of Lemma~6.1]{CKM}).  It then follows from
Stone's formula (cf. \cite[Theorem~VII.13]{RS1}) that
\begin{equation}\label{sagen2}
 \int_\R \d \omega \,  \langle \vphi, \tfrac 1 2\set{P_\omega([a,b])+P_\omega(]a,b[)}
\vphi\rangle \le \pa{b-a}.
\end{equation}
In particular, $ \int_\R \d \omega \,  \langle \vphi, P_\omega(\set{c})\vphi\rangle =0$,
and hence for any bounded interval $I$ we get
\begin{equation}\label{sagen23}
 \int_\R \d \omega \,  \langle \vphi,  P_\omega(I) \vphi\rangle \le  \abs{I}.
\end{equation}
Since $\mu$ has a bounded density $\rho$, we get
\begin{equation}\label{sagen234}
 \int_\R \d\mu(\omega) \,  \langle \vphi,  P_\omega(I) \vphi\rangle = \int_\R \d\omega \,
\rho(\omega) \langle \vphi,  P_\omega(I) \vphi\rangle \le \norm{\rho}_\infty \abs{I}.
\end{equation}

\begin{remark}
The reader may notice that \cite{CKM} has an extra factor of $\pi$ in the right hand
side of \eq{sagen234}; the difference comes from  using
 Stone's formula instead of the simple estimate  \eq{simpleineq} .  Since in Theorem~\ref{thmM}  we obtain  \eqref{minami2} as a consequence of  \eq{sagen234}, and   \eqref{minamiest} is a particular case of   \eqref{minami2},   we do not have the factor of  $\pi^2$ in the right hand side \eqref{minamiest}: the estimate is  just  $ \pa{ \rho^{(\Lambda)}_\infty \abs{I} \abs{\Lambda}}^2$.
 \end{remark}

 \subsection{Arbitrary probability distribution $\mu$}
 We consider an interval  $I= [E -\eps, E + \eps]$, $\eps >0$, and set   $z= E +i \eps$
and $R_0(z)=(H_0-z)^{-1}$.  Given $\kappa >0$,
we define   real numbers $a$ and $b$ by
\beq \label{ab}
a - i b= \tfrac \kappa {2 \eps} \la\vphi,R_0(z) \vphi\ra^{-1},
\eeq
and  note that  we always have
\begin{align}\label{b1}
\tfrac 2 \kappa  b= \frac{\tfrac 1 {\eps}
\Im\la\vphi,R_0(z)\vphi\ra}{|\la\vphi,R_0(z)\vphi\ra|^2}
= \frac{\|R_0(z)\vphi\|^2}{|\la\vphi,R_0(z)\vphi\ra|^2}
\ge 1.
\end{align}
From \eq{ab} and  \eq{rank1} we get,
\beq\label{Im}  \eps  \Im\langle \vphi, (H_\omega-z)^{-1}\vphi \rangle =\tfrac \kappa {2 } \frac b {(a +
\frac \kappa {2 \eps}\omega)^2 + b^2}.
\eeq
Proceeding as in \cite{CHK2}, and using \eq{Im},   \cite[Lemma 3.1]{CHK2} and \eq{b1}, we
get
\begin{align}\label{genav}
&\eps \int \d \mu(\omega) \,    \Im\langle \vphi, (H_\omega-z)^{-1}\vphi \rangle=\tfrac
\kappa {2 }\sum_{n\in \Z} \int_{[n\frac {2\eps }{\kappa}, (n+1)\frac {2\eps }{\kappa}[}
\d \mu(\omega) \frac b {(a +  \frac \kappa {2 \eps} \omega)^2 + b^2} \; \\
&\; \le \tfrac \kappa {2 } S_\mu \pa{\tfrac {2\eps }{\kappa}} \sum_{n\in \Z} \sup_{y \in
[0,1[ }   \frac b {(a + n + y)^2 + b^2}   \le \tfrac \kappa {2 }  \pi(1 + \tfrac 1 b)
S_\mu \pa{\tfrac {2\eps }{\kappa}}\le  \pi (1 +  \tfrac \kappa {2 }) S_\mu \pa{\tfrac
{2\eps }{\kappa}}.\notag
\end{align}

We may now get \eq{sa} in two ways.  Using the simple  inequality
 \beq \label{simpleineq}
  P_\omega (I) \le 2 \eps \,  \Im (H_\omega-z)^{-1},
 \eeq
 \eq{sa} follows immediately from \eq{genav} with
\beq
Q_\mu(\abs{I})=\inf_{\kappa >0} \pa{\pi  (2+  \kappa )S_\mu (\tfrac {1 }{\kappa}
\abs{I})}\le 3 \pi S_\mu ( \abs{I}).
\eeq
We can improve the constant slightly by using the more sophisticated inequality   given
in \cite[Eq.~(3.1)]{CHK2}, that is,
\beq
  P_\omega (I) \le
  \tfrac 4 \pi   \int_I  \d \lambda  \, \Im (H_\omega-\lambda - i \abs{I})^{-1},
 \eeq
together with \eq{genav},  getting  \eq{sa}  with 
\beq
Q_\mu(\abs{I})=\inf_{\kappa >0} \pa{4  (1+  \kappa )S_\mu (\tfrac {1 }{\kappa}
\abs{I})}\le 8 S_\mu ( \abs{I}).
\eeq

%%%%%%%%%%%%%%%%%%%%%%%%%%%%%%%%%%%%%
%%%%%%%%%%%%%%%%%%%%%%%%%%%%%%%%%%%%%
%%%%%%%%%%%%%%%%%%%%%%%%%%%%%%%%%%%%%
%%%%%%%%%%%%%%%%%%%%%%%%%%%%%%%%%%%%%
\begin{acknowledgement}
The authors thank E. Kritchevski for pointing to them the use of Stone's formula in
\eqref{sagen2}
\end{acknowledgement}

\section{An approximation lemma}  \label{secapproxlem}

\begin{lemma}\label{lemaprox} Let $F$ be a bounded, nonnegative Borel measurable function on $\R^N$, $s_1,s_2,\ldots,s_q >0$, and   $\bom=\{ \omega_j \}_{ j=1,2,\ldots,N}
$ a family of independent
random
variables, $\mu_j $ denoting  the probability distribution of the   random variable $\omega_j $. We write $\bmu=\{ \mu_j \}_{ j=1,2,\ldots,N}$, and denote the corresponding expectation by $\E_\bmu$.  Let $Q_\bmu(s):= \max_{ j=1,2,\ldots,N} Q_{\mu_j}(s)$.
Suppose there exists a constant $K>0$ such that when $\mu_j $ has compact support for all  $j=1,2,\ldots,N$
 we have
\beq\label{Funiform}
\E_\bmu\set{ F(\bom)}\le K \prod_{i=1}^q Q_\bmu(s_i).
\eeq
Then \eq{Funiform} holds for arbitrary probability distributions  $\bmu=\{ \mu_j \}_{ j=1,2,\ldots,N}$.
\end{lemma}

\begin{proof} Given a probability distribution $\mu$ and $M \in \N$, we set
$\chi_M=\chi_{[-M,M]}$ and
\beq
\mu\up{M}= c_{\mu\up{M}}\chi_M \mu, \quad \text{with}\quad  c_{\mu\up{M}}= \pa{\mu\set{[-M,M]}}^{-1}.
\eeq
Note that  $\mu\up{M}$ is a probability measure with compact support for all $M \in \N$, and \beq\label{cM}
\lim_{M\to \infty} c_{\mu\up{M}}=1.
\eeq
 Moreover, we have
\beq \label{QM}
Q_{\mu\up{M}}(s)\le c_{\mu\up{M}} Q_{\mu}(s)\quad \text{for all}\quad s > 0.
\eeq

Now, given  probability distributions   $\bmu=\{ \mu_j \}_{ j=1,2,\ldots,N}$, and $M \in \N$, we consider the  probability distributions   $\bmu\up{M}=\{ \mu_j\up{M} \}_{ j=1,2,\ldots,N}$, and set $ c_{\bmu\up{M}} = \prod_ {j=1}^N c_{\mu_j\up{M}}$.  We have
\beq\label{Funiform2}
\E_{\bmu\up{M}}\set{ F(\bom)}=  c_{\bmu\up{M}} \E_\bmu\set{\pa{\prod_ {j=1}^N \chi_M(\omega_j)} F(\bom)},
\eeq
and hence it follows from the bounded convergence theorem and \eq{cM}  that
\beq\label{Mlim}
\lim_{M\to \infty} \E_{\bmu\up{M}}\set{ F(\bom)}= \E_{\bmu}\set{ F(\bom)}.
\eeq
Since  \eq{Funiform} holds for $\bmu\up{M}$, the lemma follows from \eq{cM}, \eq{QM}, and \eq{Mlim}.
\end{proof}

%%%%%%%%%%%%%%%%%%%%%%%%%%%%%%%%%%%%%
%%%%%%%%%%%%%%%%%%%%%%%%%%%%%%%%%%%%%
%%%%%%%%%%%%%%%%%%%%%%%%%%%%%%%%%%%%%
%%%%%%%%%%%%%%%%%%%%%%%%%%%%%%%%%%%%%
%%%%%%%%%%%%%%%%%%%%%%%%%%%%%%%%%%%%%
%%%%%%%%%%%%%%%%%%%%%%%%%%%%%%%%%%%%%


\begin{thebibliography}{[CoHK2}

\bibitem[A]{A}  Aizenman, M.: {Localization at weak disorder: some
elementary bounds}. Rev. Math. Phys. {\bf 6}, 1163-1182 (1994)

\bibitem[AM]{AM}  Aizenman, M.,  Molchanov, S.:  {Localization at large
disorder and extreme energies:  an elementary derivation}.  Commun. Math.
Phys. {\bf 157}, 245-278 (1993)

\bibitem[ASFH]{ASFH}  Aizenman, M.,  Schenker, J., Friedrich, R.,
Hundertmark, D.: Finite volume fractional-moment criteria for
Anderson localization.
Commun. Math. Phys. \textbf{224}, 219-253 (2001)

\bibitem[AW]{AW}  Aizenman, M., Warzel, S.: 
 On the joint distribution of energy levels 
of random Schr\"odinger operators. Preprint 
     
\bibitem[BHS]{BHS} Bellissard, J., Hislop, P., Stolz, G.: Correlation estimates in the
Anderson model.  J. Stat. Phys.  {\bf 129},  649-662 (2007)

\bibitem[CKM]{CKM}  Carmona, R.,  Klein, A.,   Martinelli, F.: {Anderson
localization for  Bernoulli and other singular potentials}.  Commun.
Math. Phys. {\bf 108}, 41-66 (1987)

\bibitem[CoGK]{CGK}  Combes, J.M., Germinet, F., Klein, A.: {Poisson Statistics for Eigenvalues of  Continuum  Random Schr\"odinger Operators}. Preprint arXiv:0807.0455v3 [math-ph]

\bibitem[CoH]{CH} Combes,  J.M.,   Hislop, P.D.: {Localization for some
continuous, random Hamiltonians in d-dimension}. J. Funct. Anal. \textbf{124},
149-180 (1994)

\bibitem[CoHK1]{CHK} Combes, J.-M., Hislop, P. D., Klopp, F.: Regularity properties for
the density of states of random Schràdinger operators.  {\em Waves in periodic and random
media} (South Hadley, MA, 2002),  15-24, Contemp. Math., 339, 2003.

\bibitem[CoHK2]{CHK2} Combes,  J.M.,   Hislop, P.D.,  Klopp, F.:
 Optimal Wegner estimate and its application to the global continuity of the integrated
density of states for random Schr\"odinger operators.  Duke Math. J.~\textbf{140},
469-498 (2007)

\bibitem[DLS]{DLS} Delyon, F., L\'evy, Y., Souillard, B.: Anderson localization for
multidimensional systems at large disorder or large energy. Comm. Math. Phys. 100,
no. 4, 463-470 (1985)

 \bibitem[DrK]{DK} von Dreifus, H.,  Klein, A.:  {A new proof of localization in
the Anderson tight binding model}.  Commun. Math. Phys. \textbf{124},
285-299  (1989)

 \bibitem[FMSS]{FMSS} Fr\"ohlich, J.:    Martinelli, F.,  Scoppola, E.,
Spencer, T.:  {Constructive proof of localization in the Anderson tight
binding model}. Commun. Math. Phys. {\bf 101}, 21-46 (1985)



 \bibitem[FS]{FS}  Fr\"ohlich, J.,  Spencer, T.: {Absence of diffusion with
Anderson tight binding model for large disorder or low energy}. Commun.
Math. Phys. {\bf 88}, 151-184 (1983)

\bibitem[GK1]{GK1} Germinet, F.,  Klein, A.: {Bootstrap multiscale analysis and
localization in random media}.  Commun. Math. Phys. \textbf{222},
415-448 (2001)

\bibitem[GK2]{GKduke} Germinet, F.,  Klein, A.: {A characterization of the
Anderson metal-insulator transport transition}.
Duke Math. J. \textbf{124}, 309-351 (2004).


\bibitem[GK3]{GK} Germinet, F,  Klein, A.: New characterizations of
the region of complete localization for random Schr\"odinger operators.
 J. Stat. Phys.~{\bf 122}, 73-94 (2006)
 
 
 


\bibitem[GrV]{GV} Graf, G.-M., Vaghi, A.: A remark on an estimate by Minami,  Lett. Math.
Phys.  {\bf 79},  17-22 (2007)

\bibitem[H]{H}  Hundertmark, D.:  On the time-dependent approach to Anderson localization.
 Math. Nachr.  \textbf{214}, 25-38 (2000)

\bibitem[KN]{KN} Killip, R., Nakano, F.: Eigenfunction statistics in the localized
Anderson model.  Ann. Henri Poincar\'e {\bf 8},   27-36 (2007)

\bibitem[Ki]{Ki} Kirsch, W.: An invitation to random Schr\"odinger operators, Panorama et
Synth\`eses 2008.



\bibitem[KlLM]{KLM} Klein, A., Lenoble, O., M\"uller, P. : On Mott's formula for the
ac-conductivity in the Anderson model. Annals of Math.~{\bf 166}, 549-577 (2007)

\bibitem[KlM]{KM} Klein. A., Molchanov, S.: Simplicity of eigenvalues in the Anderson
model.  J. Stat. Phys.~{\bf 122}  , 95-99 (2006)

\bibitem[Kr]{Kr} Kritchevski, E.: Poisson statistics of eigenvalues in the hierarchical Anderson model. Annales Henri Poincar\'e {\bf 9}, 685-709 (2008)

\bibitem[M]{M} Minami N.: Local fluctuation of the spectrum of a multidimensional
Anderson tight binding model. Commun. Math. Phys.~{\bf 177}, 709-725 (1996)


\bibitem[N]{N}  Nakano, F.:   The repulsion between localization centers in the Anderson
model.
 J. Stat. Phys.  \textbf{123}, 803-810  (2006)

 \bibitem[RS]{RS1}  Reed, M.,  Simon, B.:  \emph{Methods of Modern
Mathematical Physics I: Functional Analysis}, revised and enlarged edition.
 Academic Press, 1980

\bibitem[S]{Si} Simon, B.:   Cyclic vectors in the Anderson model. Special issue
 dedicated to Elliott H. Lieb.  Rev. Math. Phys.  {\bf 6}, 1183-1185   (1994)

\bibitem[SW]{SW}   Simon, B., Wolff, T.:  Singular continuum spectrum under
rank one perturbations and localization for random Hamiltonians.
Commun. Pure. Appl. Math. {\bf 39}, 75-90 (1986)

\bibitem[St1]{St1} Stoiciu, M.: The statistical distribution of the zeros of random paraorthogonal polynomials on the unit circle. J. Approx. Theory~{\bf 139}, 29-64 (2006)

\bibitem[St2]{St2} Stoiciu, M.: Poisson statistics for eigenvalues: from random Schršdinger operators to random CMV matrices. Probability and mathematical physics, 465--475, CRM Proc. Lecture Notes, 42, Amer. Math. Soc., Providence, RI, 2007

\bibitem[W]{W} Wegner, F.: Bounds on the density of states in disordered systems, {Z.
Phys. B}{\bf 44} {9-15} (1981)


\end{thebibliography}
\end{document}